\documentclass[prb,twocolumn,amsmath,amssymb,prl,showpacs,superscriptaddress]{revtex4-1}
\usepackage{amsfonts, hyperref, times}
\usepackage{graphicx}

\usepackage{color}

\usepackage{lastpage}
\graphicspath{{DW_in_antiferromagnetM_V2_graphics/}{DW_in_antiferromagnetM_V2_tcache/}{DW_in_antiferromagnetM_V2_gcache/}}
\DeclareGraphicsExtensions{.pdf,.svg,.eps,.ps,.png,.jpg,.jpeg}
\graphicspath {{DW_in_antiferromagnetM_V2_graphics/}{DW_in_antiferromagnetM_V2_tcache/}{DW_in_antiferromagnetM_V2_gcache/}}
\DeclareGraphicsExtensions {.pdf,.svg,.eps,.ps,.png,.jpg,.jpeg}
\graphicspath {{DW_in_antiferromagnetM_V1_graphics/}{DW_in_antiferromagnetM_V1_tcache/}{DW_in_antiferromagnetM_V1_gcache/}}
\DeclareGraphicsExtensions {.pdf,.svg,.eps,.ps,.png,.jpg,.jpeg}

\begin{document}

\title{Narrow-band tunable THz detector in antiferromagnets via N\'eel spin-orbit torque and spin-transfer torque}

\author{O. Gomonay}
\affiliation{Institut f\"ur Physik, Johannes Gutenberg Universit\"at Mainz, D-55099 Mainz, Germany}
\affiliation{National Technical University of Ukraine ``KPI'', 03056, Kyiv,
	Ukraine}
\author{T. Jungwirth}
\affiliation{Institute of Physics ASCR, v.v.i., Cukrovarnicka 10, 162 53 Praha 6 Czech Republic}
\affiliation{School of Physics and Astronomy, University of Nottingham, Nottingham NG7 2RD, United Kingdom}
\author{J. Sinova}
\affiliation{Institut f\"ur Physik, Johannes Gutenberg Universit\"at Mainz, D-55099 Mainz, Germany}
\affiliation{Institute of Physics ASCR, v.v.i., Cukrovarnicka 10, 162 53 Praha 6 Czech Republic}

\begin{abstract}
We study dynamics of antiferromagnets induced by simultaneous application of dc spin current and ac charge current, motivated by the requirement of all-electrically controlled devices in THz gap (0.1-30 THz). We show that ac electric current, via N\'eel spin orbit torques, can lock the phase of a steady rotating N\'eel vector whose precession is controlled by a dc spin current. In the phase-locking regime the frequency of the incoming ac signal coincides with the frequency of autooscillations which for typical antiferromagnets fall into the THz range. The frequency of autooscillations is proportional to the precession-induced tilting of the magnetic sublattices related to the so-called dynamical magnetization. We show how the incoming ac signal can be detected from the measurement of the dc-current dependencies of the constant dynamical magnetization.  We formulate the conditions of phase-locking based on relations between parameters of an antiferromagnet and the characteristics of the incoming signal (frequency, amplitude, bandwidth, duration). We also show that the rotating N\'eel vector can generate ac electrical current via inverse N\'eel spin-orbit torque. Hence, antiferromagnets driven by dc spin current can  be used as  tunable detectors and emitters of narrow-band signals operating in the THz range. 
\end{abstract}

\maketitle
The information processing speed in modern communication systems is defined by the signal frequency, which can be either below 0.1 THz for electrical sources or above 30 THz for the laser-based sources. However, development of the relevant technologies in the range between 0.1-30 THz --  also referred to as a Teraherz gap  -- started only recently \cite{Dhillon2017}. Therefore, it is important to find new materials and physical phenomena suitable to fill in the Teraherz gap.    

From this point of view, antiferromagnets (AFs) with high N\'eel temperature are promising materials, as their resonance frequencies fall into the THz range and they can be effectively manipulated by spin and charge currents. Moreover, it was previously reported \cite{Gomonay2010, Cheng2015, Cheng2016, Khymyn2017} that a dc spin-polarized current can induce steady precession of the N\'eel vector at the THz frequency which scales with the current density. Hence, an antiferromagnet can be considered as a tunable current controlled autooscillator, which is the main component of standard phase-controlling devices (detectors, amplifiers etc). However, application of an antiferromagnet as a detector or emitter of Teraherz radiation is often impeded by its weak coupling with external magnetic fields. In this work we theoretically show that the detection of THz radiation can be achieved by phase-locking phenomena in the AFs that exhibit the effect of the 
recently predicted \cite{Zelezny2014} and observed \cite{Wadley2016,Grzybowski2017,Bodnar2017}
N\'eel spin orbit torque (NSOT).

Autooscillations are undamped oscillations, whose amplitude and frequency are  independent of the initial conditions and determined by the properties of the system itself. They can be generated in AFs by spin-transfer or spin-orbit torques of anti-damping character. These torques, typically induced by dc current, fully compensate the internal damping and thus sustain steady rotation of the N\'eel vector. The frequency of these induced autooscillations is proportional to the torque strength. Similar to all autooscillating systems,  AFs in an autooscillating regime should show the effect of phase-locking (enforced synchronization), where the rotation of the N\'eel vector assumes the frequency of an external ac signal. Here such a signal is required to have a  field-like torque character relative to the N\'eel vector,  
which in our case  is provided by the NSOT.

 The NSOT can arise in antiferromagnets with two magnetic sublattices  which form inversion partners, while the local inversion symmetry of the magnetic atoms is broken. In these antiferromagnets the applied uniform electric field/charge current  produces locally a non-equilibrium spin accumulation which alternates in sign between the different magnetic sublattices, and results in a staggered spin-orbit field that strongly couples to the N\'eel order parameter. This staggered field is an antiferromagnetic analogue of the Zeeman magnetic field in ferromagnetic materials and therefore has a field-like character. The field-like NSOT can be created by an ac electric current or by the electrical component of an electromagnetic wave.\cite{Olejnik2017a}  
 
In this Letter we study dynamics of the N\'eel vector generated by simultaneously applying  a dc spin-polarized current and an ac charge current, which can lead to a phase-locking regime. We show that in the region of phase-locking the dependence of the precession frequency of the autooscillations vs dc current has a horizontal segment when the autooscillations and the external signal are synchronized. The phase-locking interval is influenced by the amplitude and bandwidth of the incoming ac signal. The phase-locking effect is also sensitive to the polarization of the incoming signal and allows to discern between two opposite circular  modes. We propose a practical realization of the AF-based phase-locking detector by calculating the relation between the precession frequency and the observable -- constant dynamical magnetization of an AF. We further demonstrate  that in the state of spin-current induced precession the AF can also generate an ac current through the \emph{inverse NSOT}, the effect which we predict based on Onsager reciprocity relations. Hence, the (inverse) NSOT in combination with spin current can be potentially used to create tunable detectors (emitters) of narrow-band signals that can operate in the THz gap range.

{\it Model --} We consider a collinear compensated AF whose crystalline symmetry allows for field-like NSOT. \cite{Zelezny2014,Zelezny2017} To aid the practical realization of the system, we focus on the materials CuMnAs and Mn$_2$Au, in which NSOT has been recently experimentally demonstrated \cite{Wadley2016, Bodnar2017,Meinert2017}
Their tetragonal easy plane magnetic structure and  parameters are collected in Table~\ref{Tab_data}. The magnetic structure of the AF is represented by two vectors of sublattice magnetizations, $\mathbf{M}_1$ and $\mathbf{M}_2$, which are antiparallel and fully compensate each other in the equilibrium state. While moving, these vectors are slightly tilted forming nonzero magnetization $\mathbf{m}=\mathbf{M}_1+\mathbf{M}_2$.

 However, due to the strong exchange coupling between the magnetic sublattices (parametrized with the constant $H_\mathrm{ex}$), the magnetization $\mathbf{m}$ is small and the state of the antiferromagnet is fully described by the  N\'eel vector (or staggered magnetization) $\mathbf{n}=\mathbf{M}_1-\mathbf{M}_2$ whose magnitude $|\mathbf{n}|=2M_s$ is fixed well below the N\'eel temperature. 
 
\begin{figure}[h]
	\centering
\includegraphics[width=1\columnwidth]{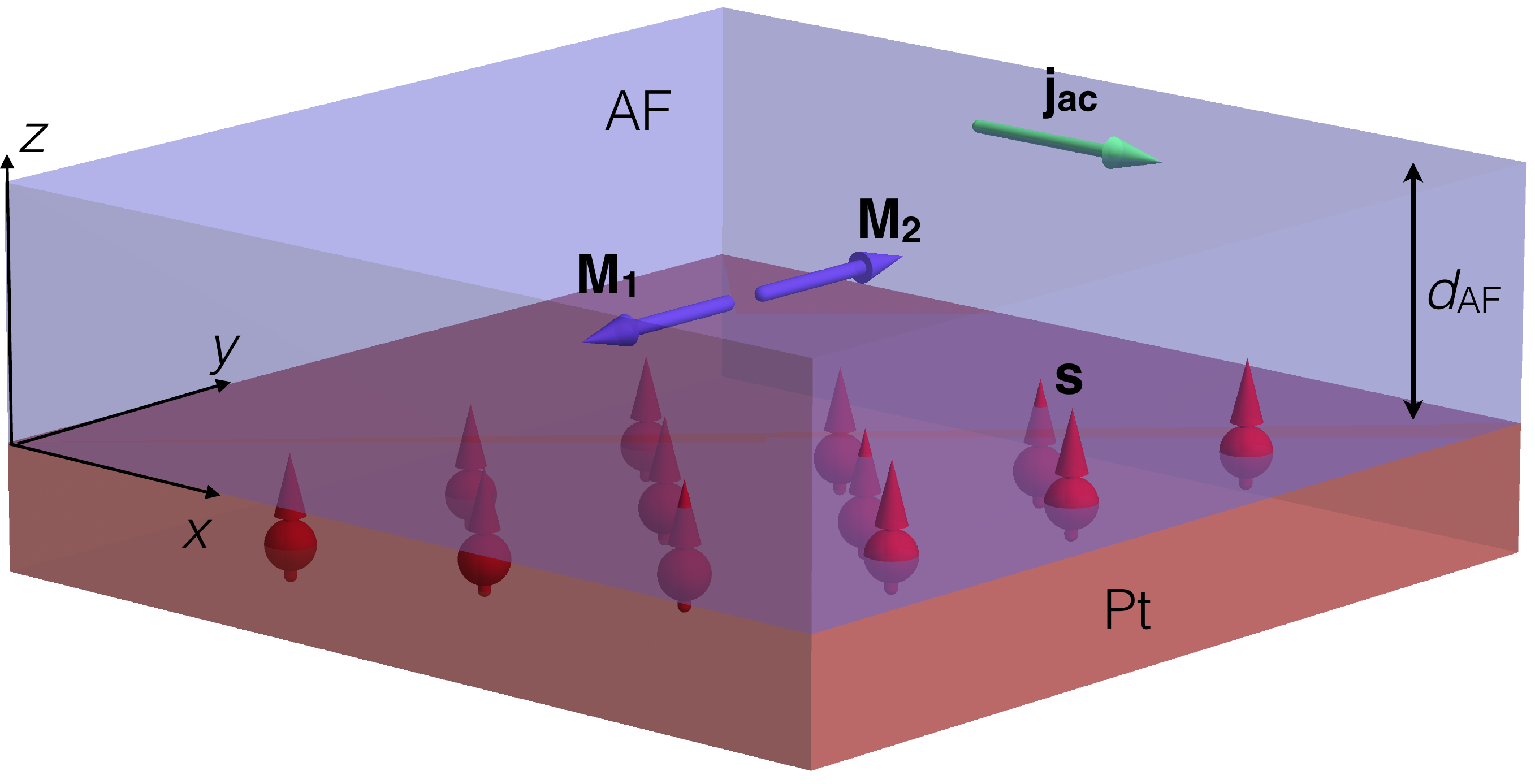}
	\caption{(Color online) Scheme of a bilayer system of an antiferromagnet (AF) and a heavy metal (Pt) for detection of THz signal.  The dc spin polarized current with spin polarization $\mathbf{s}$ is created due to the inverse spin-Hall effect in a heavy metal layer, the incoming signal is created by an ac charge current with the current density $\mathbf{j}_\mathrm{ac}$ within an AF layer. The spin polarized current induces rotation of the staggered magnetization $\mathbf{M}_1\uparrow\downarrow\mathbf{M}_2$ within the film plane. The frequency of the rotations can be locked by the incoming ac signal.}
	\label{fig_scheme}
	\vskip -0.5 cm
\end{figure}

 The dynamics of the N\'eel vector is driven by two external torques: i) anti-damping-like N\'eel torque \cite{Gomonay2008,Zelezny2014} $\propto \mathbf{n}\times\mathbf{s}\times\mathbf{n}$, which emerges from the \emph{dc} spin polarized current 
 with spin polarization $\mathbf{s}$, $|\mathbf{s}|=1$ (Fig.~\ref{fig_scheme}));
 and ii) field-like NSOT \cite{Zelezny2014} $\propto \mathbf{n}\times\hat{z}\times\mathbf{j}_\mathrm{ac}$, created by an \emph{ac} charge current with density $\mathbf{j}_\mathrm{ac}$. In the macrospin approximation, valid for a single domain, the equation of motion for the N\'eel vector is:\cite{Gomonay2008, Cheng2014c, Gomonay2016}
\begin{eqnarray}\label{eq_motion_antiferromagnet_initial}
&&\mathbf{n}\times(\ddot{\mathbf{n}}+2\alpha_G\gamma H_\mathrm{ex}\dot{\mathbf{n}}-2\gamma^2H_\mathrm{ex}M_s\mathbf{H}_\mathbf{n})\\
&&=\gamma H_\mathrm{ex}\mathbf{n}\times(\gamma H_\mathrm{dc}\mathbf{s}\times\mathbf{n}+2\lambda_\mathrm{NSOT} M_s\mathbf{j}_\mathrm{ac}\times\hat{z}),\nonumber
\end{eqnarray}
where $\alpha_G$ is the Gilbert damping constant, $\gamma$ is the gyromagnetic ratio, and $\mathbf{H}_\mathbf{n}=-\partial w_\mathrm{an}/\partial \mathbf{n}$ is the  internal effective field defined by the profile of magnetic anisotropy energy $w_\mathrm{an}$. For convenience, we characterise the effective density of spin-polarized current with the value $H_\mathrm{dc}$ which has the dimensionality of the magnetic field. In the particular case when the spin current is generated by the spin Hall effect (see Fig.~\ref{fig_scheme}), $H_\mathrm{dc}=\hbar \varepsilon \theta_\mathrm{H}j_\mathrm{dc}/(2ed_\mathrm{AF}M_s)$, where $\hbar$ is the Planck constant, $d_\mathrm{AF}$ is the thickness of the film, $0<\varepsilon\le1$ is the spin-polarization efficiency, $\theta_{H}$ is the bulk Hall angle, $e$ is an  electron charge, and $j_\mathrm{dc}$ is the  dc current density in the heavy metal electrode. Although for the spin Hall effect the polarization is typically in-plane, a side structure (different from the one show in Fig.~\ref{fig_scheme}) would be necessary. We use the spin Hall effect here just for the purposes of estimating the necessary order of magnitude of the currents.

The last term on the right hand side of Eq.~(\ref{eq_motion_antiferromagnet_initial}) describes the effect of the field-like NSOT assuming an ac steady current.  Note that this expression can be  applied for an ac currents as long as the oscillation period (picoseconds) is much larger than the electron relaxation time (femptoseconds).

We demonstrate next that an AF is a perfect candidate for the phase-sensing detector as it possesses all necessary features: i) has a tunable autooscillating regime; ii) efficiently couples with the electrical ac signal; and iii) creates a measurable response.  
 
 \begin{figure}[h]
	\centering
\includegraphics[width=1\columnwidth]{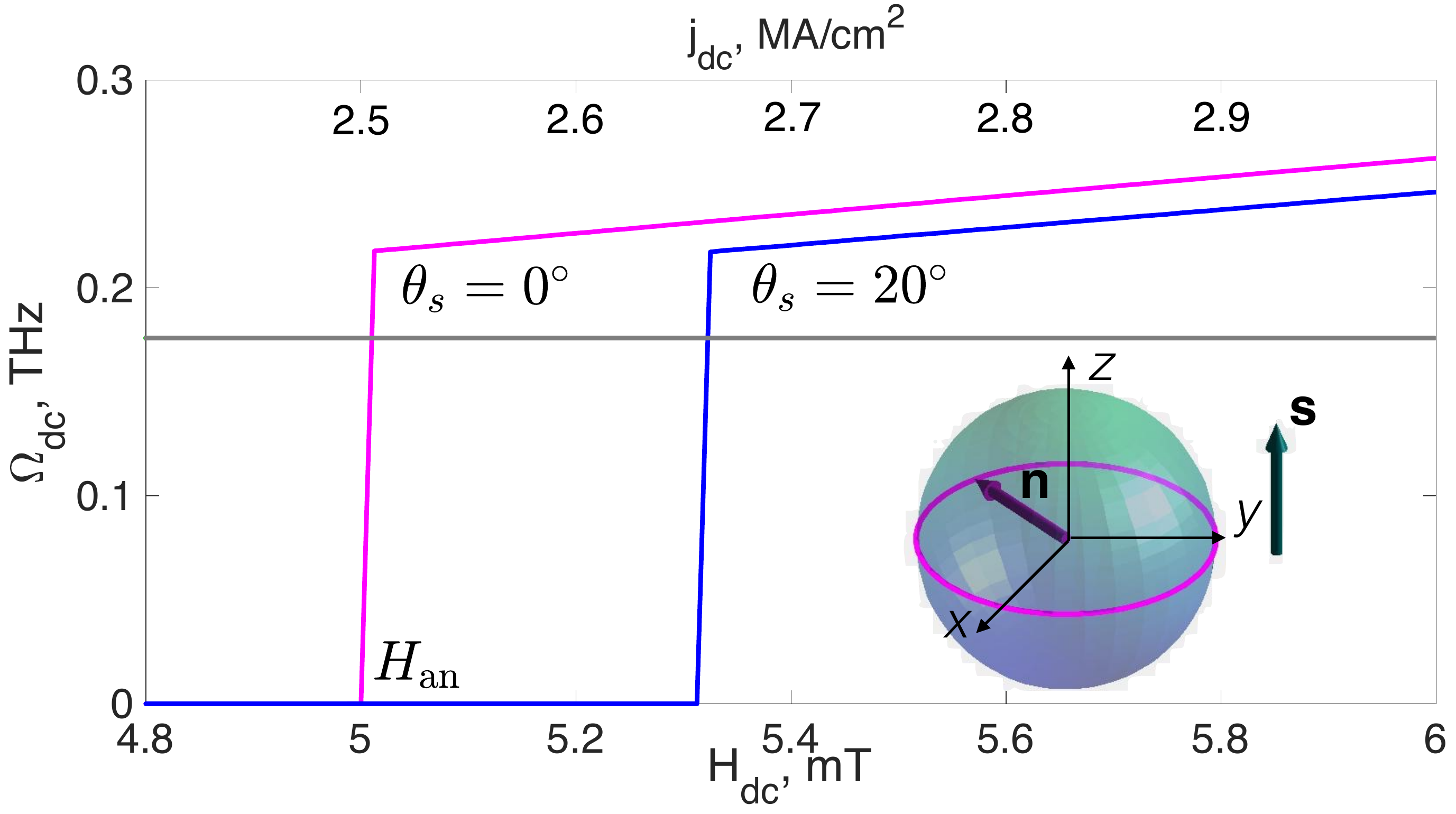}
	\caption[Spin-current induced rotation of antiferromagnet moment]{(Color online) Average frequency of spin-current induced rotation of the N\'eel vector $\mathbf{n}$ as a function of dc current density calculated for CuMnAs from Eq.~(\ref{eq_motion_antiferromagnet_initial}). Above the threshold the frequency grows linearly  with the current value. 
The value of the threshold current is proportional to the in-plane magnetic anisotropy, $H_\mathrm{an}$,  and depends on the angle $\theta_\mathrm{s}$ between spin current and hard axis. 
Minimal threshold ($=H\mathrm{an}$) is observed when $\mathbf{s}$ is parallel to the hard axis ($\theta_\mathrm{s}=0^\circ$, magenta). 
Horizontal line shows the frequency $\omega_\mathrm{AFR}/2$.  Field-current conversion $H_\mathrm{dc}/j_\mathrm{dc}=1$ mT/(MA/cm$^2$) corresponds to spin-pumping via spin Hall effect with the Hall angle $\theta_\mathrm{H}=0.1$\cite{Sinova2015} into the sample with thickness $d_\mathrm{AF}=1$~nm.}
	\label{fig_frequency_vs_dc_current_stability}
\end{figure}

{\it Antiferromagnet as a spin-torque oscillator --}
To study the autooscillation regime of the antiferromagnet we solve Eq.~(\ref{eq_motion_antiferromagnet_initial}) in the presence of a dc spin current only ($\mathbf{j}_\mathrm{ac}=0$) for different orientations of the spin polarization $\mathbf{s}$ and arbitrary initial conditions for the N\'eel vector $\mathbf{n}$. In agreement with  previous studies, \cite{Gomonay2010, Cheng2015} the spin-polarized current induces steady precession of the N\'eel vector within the plane perpendicular to $\mathbf{s}$ (inset in Fig.~\ref{fig_frequency_vs_dc_current_stability}). This state can be achieved for any orientation of the spin-current polarization and from any initial state of an AF. However, to generate the auto-oscillations it is necessary to overcome the threshold ("hard" generation) $H_\mathrm{dc}\ge H^\mathrm{thr}_\mathrm{dc}\equiv H_\mathrm{an}/s_z$ whose value is sensitive to the orientation of $\mathbf{s}$ with respect to the easy (hard) direction. This behaviour is illustrated in Fig.~\ref{fig_frequency_vs_dc_current_stability}, which shows  the dependence of the average frequency of steady rotations, $\Omega_\mathrm{dc}$, vs the current density $H_\mathrm{dc}$ for two different orientations of $\mathbf{s}$. The minimal threshold $H^\mathrm{thr}_\mathrm{dc}=H_\mathrm{an}$ is obtained when $\mathbf{s}$ is parallel to the hard axis.

Above the threshold, the average frequency $\Omega_\mathrm{dc}$  is defined from the balance of the spin-pumping ($\gamma H_\mathrm{dc}\mathbf{s}\times\mathbf{n}$) and internal damping ($2\alpha_G\dot{\mathbf{n}}$). It grows linearly with the current value, $\Omega_\mathrm{dc}=\gamma H_\mathrm{dc}/2\alpha_G$. The absolute value of $\Omega_\mathrm{dc}$ is comparable with the AFR frequency $\omega_\mathrm{AFR}\equiv2\gamma\sqrt{H_\mathrm{an}H_\mathrm{ex}}$ of the lowest mode and thus falls into the THz range.

{\it Phase locking and signal detection --}
 In the auto-oscillation regime the components of the N\'eel vector oscillate with the frequency $\Omega_\mathrm{dc}$ and thus produce an ac ``reference signal''. These oscillations can be locked by the incoming ac signal if the difference between the two frequencies is sufficiently small.
 
To illustrate the phase-locking effect in the auto-oscillating AF  we consider a geometry in which the dc spin current is polarized along the hard axis, $\mathbf{s}\|\hat{z}$, corresponding to the minimal threshold current (Fig. \ref{fig_scheme}). We further assume that the incoming signal with frequency $\omega_\mathrm{ac}$ creates a charge current $\mathbf{j}_\mathrm{ac}$ within the easy plane. In this case the magnetic dynamics are described by the single variable $\varphi$ which defines the orientation of the N\'eel vector in the easy plane.  The equation of motion given by Eq.~(\ref{eq_motion_antiferromagnet_initial})  can be re-written as
\begin{eqnarray}\label{eq_simplified}
&&\frac{1}{\gamma H_\mathrm{ex}}\ddot{\varphi}+2 \alpha_\mathrm{eff}\dot{\varphi}+\gamma H_\mathrm{an}\sin(4\varphi)\\
&&=\gamma H_\mathrm{dc}-\lambda_\mathrm{NSOT}\left(j^x_\mathrm{ac}\cos\varphi+j^y_\mathrm{ac}\sin\varphi\right),\nonumber
\end{eqnarray}
where the constant $\alpha_\mathrm{eff}$ includes contributions from both Gilbert damping and field-like NSOT,  and $H_\mathrm{an}$ is the in-plane anisotropy field. If not specified, the ac signal is monochromatic and linearly polarized (created by a linearly polarized electromagnetic wave with the electrical component $\mathbf{E}\|x$), $j^x_\mathrm{ac}=j^{(0)}_\mathrm{ac}\cos\omega_\mathrm{ac}t, j^y_\mathrm{ac}=0$.

Above the threshold dc current, {$H_\mathrm{dc}>H^\mathrm{thr}_\mathrm{dc}$}, Eq.~(\ref{eq_simplified}) can be approximately solved by separating fast and slowly varying terms. For this we assume that the N\'eel vector rotates with the average frequency $\Omega$ close to $\omega_\mathrm{ac}$, so that $\varphi=\Omega t+\psi(t)+\varphi_0$, where $\psi(t)$ is a fast variable, and $\varphi_0$ is a constant phase shift between the internal oscillations and the external force. After averaging Eq.~(\ref{eq_simplified})   over the time interval $T\gg 2\pi/\omega_\mathrm{ac}$ (denoted below with symbol $\langle\ldots\rangle$), we split Eq.~(\ref{eq_simplified}) into two related equations for $\Omega$ and $\psi$:
\begin{equation}\label{eq_4_average}
\Omega=\Omega_\mathrm{dc}
+\frac{1}{2}\Delta\omega\langle\cos\left[\left(\Omega-\omega_\mathrm{ac}\right) t+\psi(t)+\varphi_0\right]\rangle,
\end{equation}
	\begin{equation}
\ddot{\psi}+2 \alpha_\mathrm{eff}\gamma H_\mathrm{ex}\dot{\psi}=F(t),
\label{eq_fast}
\end{equation}
\noindent where $\Omega_\mathrm{dc}=\gamma H_\mathrm{dc}/2\alpha_\mathrm{eff}$ is the average frequency in the absence of the ac signal, $\Delta\omega\equiv \lambda_\mathrm{NSOT} j^{(0)}_\mathrm{ac}/\alpha_\mathrm{eff}$, and $F(t)$ is a function whose spectrum consists of the frequencies $\Omega\mp\omega_\mathrm{ac}$ and higher harmonics of $\Omega$.

Let us assume that $\Omega=\omega_\mathrm{ac}$. In this case $F(t)$ contains only higher harmonics $2\Omega, 4\Omega,\ldots$, whose contribution into $\psi(t)$ is weakened due to small factor $ \propto1/\Omega^2$. Hence, $\psi\ll1 $ and can be neglected. Equation~(\ref{eq_4_average}) is then reduced to an equality
$\Omega=\Omega_\mathrm{dc}+{1\over 2}\Delta\omega\cos\varphi_0$  which is consistent with $\Omega=\omega_\mathrm{ac}$ if  $|\Omega_\mathrm{dc}-\omega_\mathrm{ac}|\le\Delta\omega/2$. This corresponds to the phase-locking regime where the N\'eel vector rotates with the frequency of the input ac signal.  

For $|\Omega_\mathrm{dc}-\omega_\mathrm{ac}|>\Delta\omega/2$, the autooscillations are detuned from the external signal,  $\Omega\ne \omega_\mathrm{ac}$.  In this range $F(t)$ contains a slowly varying harmonic at a frequency $(\Omega-\omega_\mathrm{ac})$  and  $\psi(t)\approx\psi_\mathrm{amp}\cos\left[\left(\Omega-\omega_\mathrm{ac}\right) t\right]$, where the amplitude of oscillations, 
\begin{equation}\label{eq_psi_approx}
\psi_\mathrm{amp}\propto\frac{1}{|\Omega-\omega_\mathrm{ac}|\sqrt{(\Omega-\omega_\mathrm{ac})^2+4\gamma^2\alpha_\mathrm{eff}^2H_\mathrm{ex}^2}},
\end{equation}
\noindent can be large enough to reduce the time-average in Eq.~(\ref{eq_4_average}) to zero. In this case the frequency of autooscillations $\Omega\approx\Omega_\mathrm{dc}$ is detuned from the ac signal and is defined only  by the internal properties of the system.

\begin{figure}[h]
 	\centering
 	\includegraphics[width=0.9\linewidth]{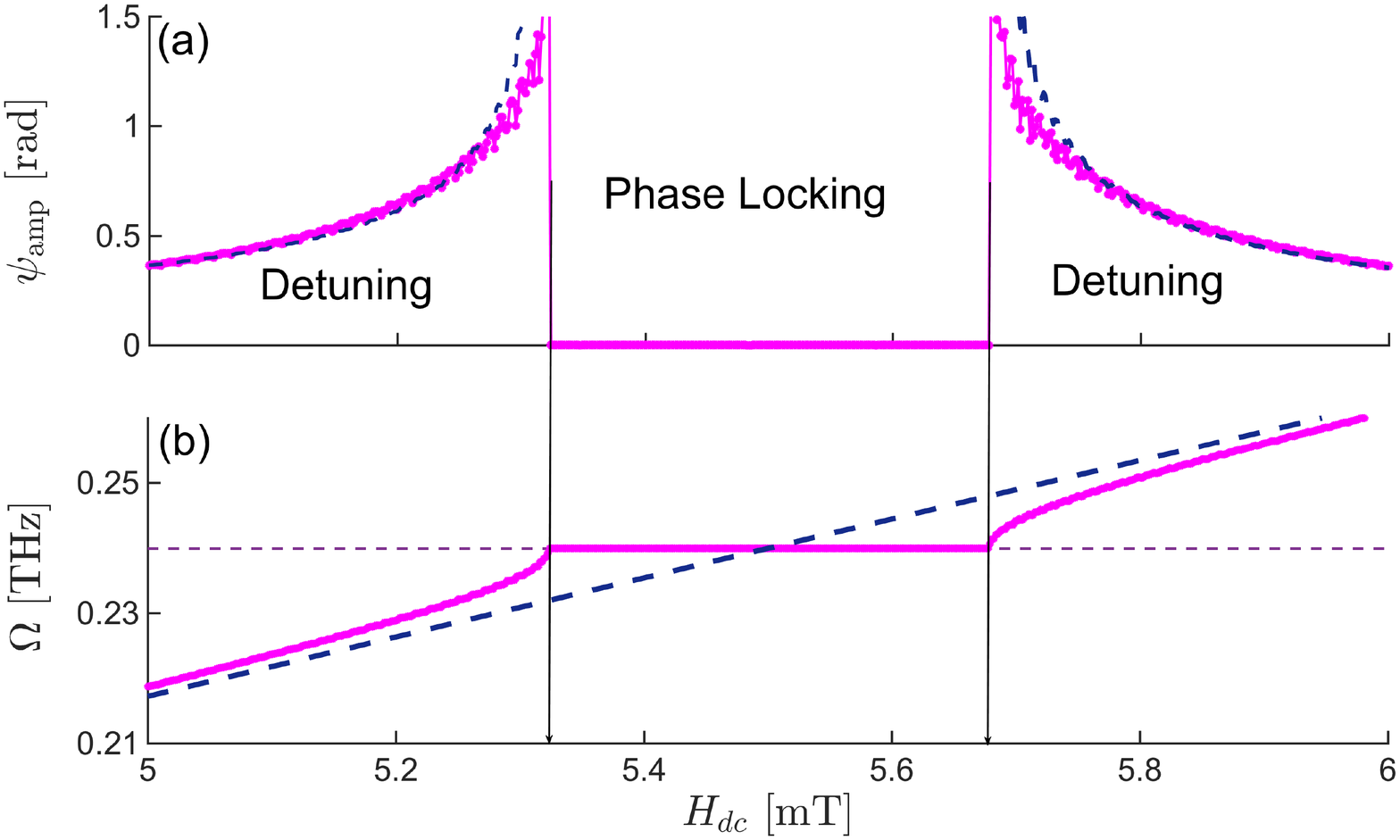}
 	\caption{(Color online) Illustration of phase locking effect: The purple (solid) lines show (a) the amplitude $\psi_\mathrm{amp}$ of the lowest harmonic at frequency $(\Omega-\omega_\mathrm{ac})$  and (b) the autooscillation frequency $\Omega$ as a function of dc current, ${H}_\mathrm{dc}$, calculated from Eq.~(\ref{eq_motion_antiferromagnet_initial}) for CuMnAs at $\omega_\mathrm{ac}=0.24$~THz, and $j_\mathrm{ac}^{(0)}=5$~MA/cm$^2$. Dashed-blue lines show the approximate solution for (a) $\psi_\mathrm{amp}$ according to  (\ref{eq_psi_approx})  and for (b) the authooscillation frequency $\Omega_\mathrm{dc}$  in the absence of an ac signal (b). 
}
 	\label{fig_phase_locking_double}
 \end{figure}

\begin{figure}[h]
	\centering
\includegraphics[width=0.9\linewidth]{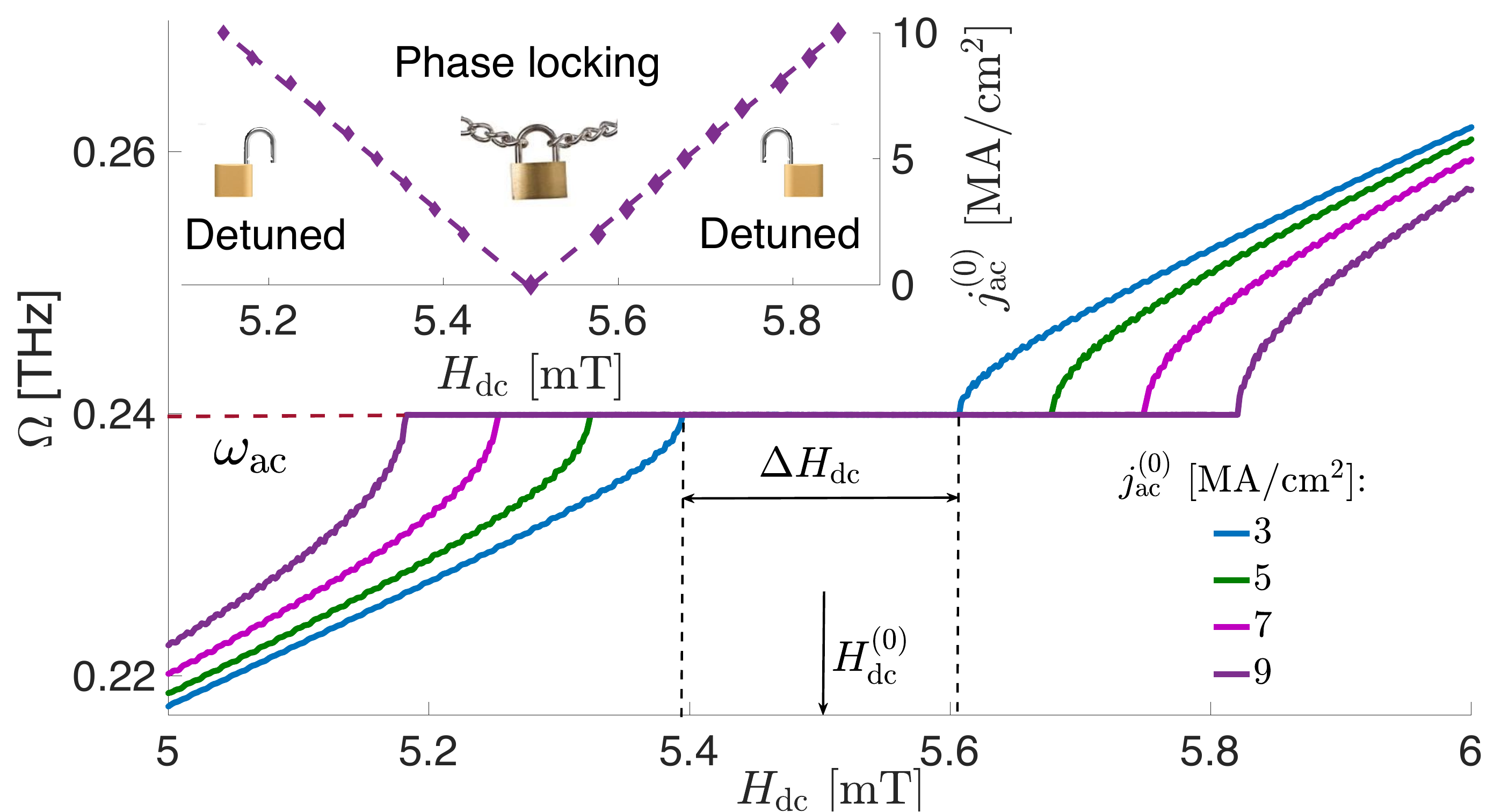}
	\caption{(Color online) Frequency of autooscillations $\Omega$ as a function of dc current calculated for CuMnAs from Eq.~(\ref{eq_motion_antiferromagnet_initial}) for different amplitudes of ac signal. Frequency of ac signal $\omega_\mathrm{ac}=0.24$~THz. Vertical dashed lines outline the phase-locking interval $\Delta H_\mathrm{dc}$ for $j_\mathrm{ac}^{(0)}=3$~MA/cm$^2$. Inset: Phase-locking region in a parameter $j_\mathrm{ac}^{(0)}-H_\mathrm{dc}$ space  (Arnold tongue): markers -- numeric simulation, dashed lines -- theoretical prediction.}
	\label{fig_magnetization_vs_dc_current}
\end{figure}

This  picture is verified by the numerical simulations of the dynamics based on Eq.~(\ref{eq_motion_antiferromagnet_initial}), as shown in Figs.~\ref{fig_phase_locking_double} and \ref{fig_magnetization_vs_dc_current}. 
In Fig.~\ref{fig_phase_locking_double} (a) and (b) we compare the behaviour of the average frequency ($\Omega$) corresponding to the slow variable and  the amplitude of the fast variable ($\psi_\mathrm{amp}$) as a function of $H_\mathrm{dc}$.  
The center of the phase locking regime is located at $H_\mathrm{dc}=H^{(0)}_\mathrm{dc}\equiv2\alpha_\mathrm{eff}\omega_\mathrm{ac}/\gamma$, defined by the signal frequency, and the  width $\Delta H_\mathrm{dc}$ correlates well with the predicted relation $\Delta H_\mathrm{dc}\equiv2\alpha_\mathrm{eff}\Delta \omega/\gamma=\lambda_\mathrm{NSOT}j^{(0)}_\mathrm{ac}/(2\gamma)$.
In this region $\psi_\mathrm{amp}=0$ (phase locking) and $\Omega=\omega_\mathrm{ac}$.

 At the borders of the phase-locking interval the phase grows up to $\psi_\mathrm{amp}\propto\pi/2$ which results in detuning of autooscillation from the incoming signal. Far from the phase-locking region $\Omega$ coincides with the internal autooscillation frequency $\Omega_\mathrm{dc}$,  corresponding to the dashed-line in Fig.~\ref{fig_phase_locking_double}b.

 Thus, due to the phase locking effect, the presence of an ac signal can be detected by measuring  the dependence of the autooscillation frequency vs dc current. Moreover, it is also possible to determine the amplitude of the ac signal from the $\Omega(H_\mathrm{dc})$ curve, since the width of the phase locking region is linear in the current amplitude, $j^{(0)}_\mathrm{ac}=2\gamma|H_\mathrm{dc}-H^{(0)}_\mathrm{dc}|/\lambda_\mathrm{NSOT}$, as shown in the inset in Fig.~\ref{fig_magnetization_vs_dc_current}.
 
\begin{figure}[h]
	\centering
\includegraphics[width=0.95\linewidth]{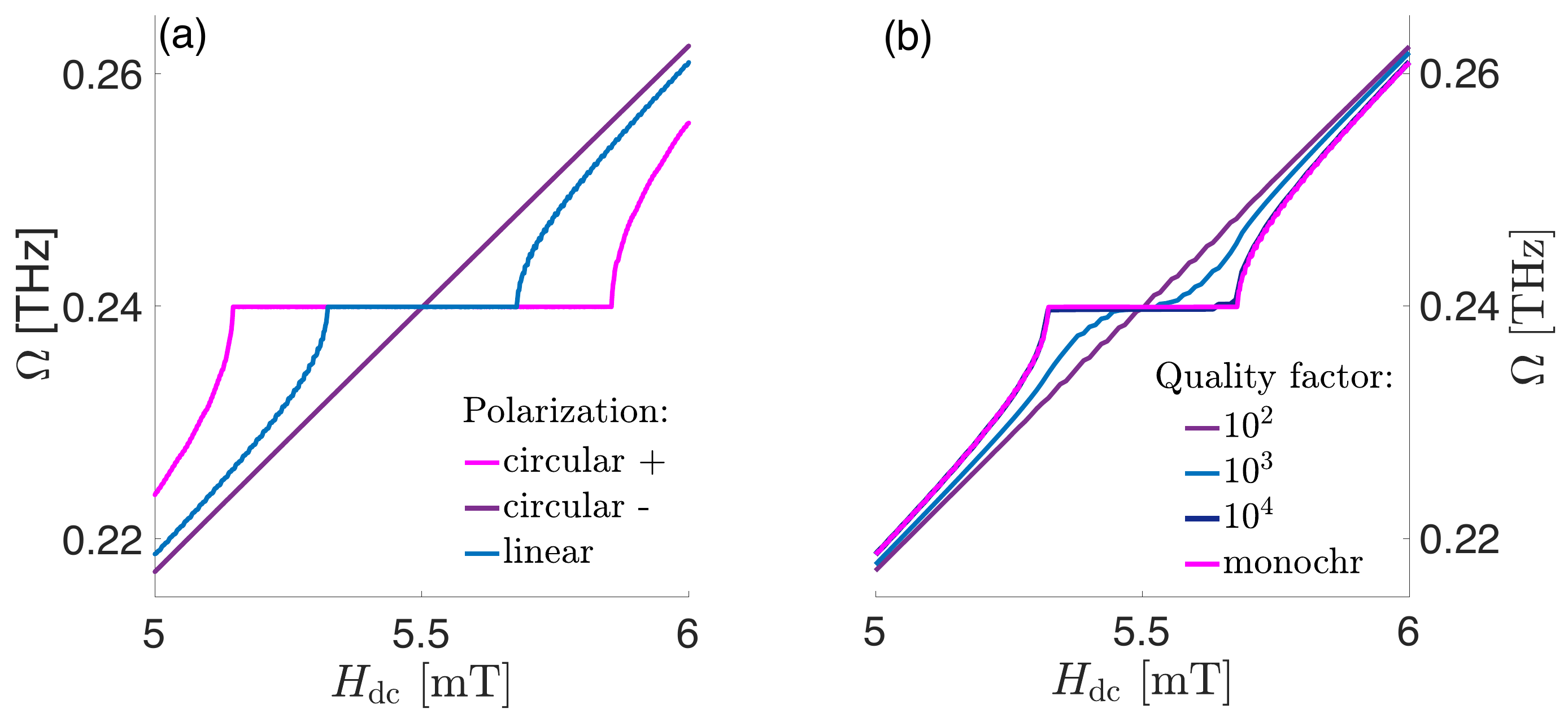}
	\caption{(Color online) Effect of (a)  polarization and (b) nonmonochromacity on phase locking. The frequency of autooscillations $\Omega$ as a function of dc current is calculated from Eq.~(\ref{eq_motion_antiferromagnet_initial}) for CuMnAs with $\omega_\mathrm{ac}=0.24$~THz, and $j_\mathrm{ac}^{(0)}=5$~MA/cm$^2$.}
	\label{fidifferent-polarizations}
	\vskip -0.25 cm
\end{figure}

The phase-locking effect allows not only to detect the signal but also to determine its polarization in the case when the ac signal is a circularly polarized electromagnetic wave. To illustrate this fact we study dynamics of the N\'eel vector assuming that 
$j^x_\mathrm{ac}=j^{(0)}_\mathrm{ac}\cos\omega_\mathrm{ac}t, j^y_\mathrm{ac}=\pm j^{(0)}_\mathrm{ac}\sin\omega_\mathrm{ac}t$. The two signs ($\pm$) correspond to counterclockwise/clockwise rotation of vector  $\mathbf{j}_\mathrm{ac}$ within the easy plane. The results are summarized in Fig.~\ref{fidifferent-polarizations}a,  which shows the dependence of $\Omega(H_\mathrm{dc})$ for counterclockwise (magenta) and clockwise (blue) polarizations. Phase locking appears only for one type of polarization, when both vectors, $\mathbf{n}$ and $\mathbf{j}_\mathrm{ac}$ rotate in the same direction. 
For comparison we also show the results for the linearly polarised current calculated for the same $j^{(0)}_\mathrm{ac}$ (green). In this case the phase locking effect  appears for any orientation of $\mathbf{j}_\mathrm{ac}$, however, the phase-locking interval is two times smaller than in the case of the circularly polarized signal.

Up to now we considered the ideal case of a monochromatic ac signal and neglected possible fluctuations of the driving dc current.
To make our results more realistic we study also the dynamics of the antiferromagnetic autooscillator induced by ac signals with a gaussian frequency distribution with   bandwidth $\Delta \omega_\mathrm{sign}$.  We assume that the signal duration $T_\mathrm{sign}\ge 2\pi/\Delta \omega_\mathrm{sign}$ and that it does not  contribute to the bandwidth. The frequency of autooscillations is then obtained by averaging over the time $T_\mathrm{sign}$. 
Typical $\Omega(H_\mathrm{dc})$ dependencies  calculated for signals with  different values of the quality factor $\omega_\mathrm{ac}/\Delta \omega_\mathrm{sign}$ are shown in Fig.~\ref{fidifferent-polarizations}b. The curves with high  quality factor (down  to $10^3$) show the wide horizontal segment, which corresponds to the phase locking at the central frequency $\omega_\mathrm{ac}$. However, as the quality factor diminishes, additional locking at the side-band frequency is also possible. In this case $\Omega(H_\mathrm{dc})$ can have a different spin current dependence of the frequencies (green line in Fig.~\ref{fidifferent-polarizations}b). Decreasing the quality factor increases the steepness of the curve and the $\Omega(H_\mathrm{dc})$ dependence is indistinguishable from the linear one (red and blue lines). So, reliable detection via phase-locking is possible for narrowband signals with $\Delta \omega_\mathrm{sign}\le 10^{-3}\omega_\mathrm{ac}$. 

Phase locking effects can also be smeared by fluctuations of the dc current whose amplitude defines the bandwidth of autooscillations in the absence of an ac signal. Optimally, this bandwidth should be kept of the order or below the bandwidth of the detected signal. Hence, signal-to-noise ratio for the dc current should be above 10$^3$. It should be also noted that, as the autooscillation frequency is defined mainly by the dc current, fluctuations of the magnetic constants (anisotropy and exchange fields), have minor influence on the phase locking conditions. The morphology of the sample (domains and grains) is also unimportant because in the autooscillation regime the information about the initial state is lost at the timescale of the magnetic relaxation time $1/(\alpha_\mathrm{eff}\gamma H_\mathrm{ex})$, which is supposed to be much smaller than the signal duration $T_\mathrm{sign}$. 

Our suggested phase-locking detection is based on  frequency measurements which  can be challenging. However, rotation of the N\'eel vector creates a nonzero magnetization $\mathbf{m}=\mathbf{n}\times\dot{\mathbf{n}}/(2\gamma H_\mathrm{ex}M_s)$ \cite{Baryakhtar1980}. In the autooscillation regime  this magnetization is \emph{constant}, parallel to the rotational axis,  and its value is proportional to the autooscillation frequency, $m_z=2\Omega M_s/\gamma H_\mathrm{ex}$. For CuMnAs at an autooscillation frequency of 0.25 THz the magnetization can reach 16 A/cm and create a magnetic field of up to 2 mT (see Table~\ref{Tab_data}). The measurement of such magnetization needs high-sensitive measuring tools. However, the technique based on NV centers\cite{Jensen2014} gives an accuracy up to 20 $\mu$T which makes the detection of magnetization  plausible.

{\it Inverse N\'eel spin-orbit torque effect and generation of THz signal --}
Next we demonstrate that the same coupling mechanism that allowed us to construct a THz detector should allow for the generation of an ac current by the oscillating N\'eel vector. The physics of this phenomenon, which we call inverse NSOT effect, is similar to the physics of the spin-galvanic effect \cite{Ganichev2002}. In  metallic and semiconducting magnets the distribution of spin density of free electrons is correlated with the orientation of the localized magnetic moments. Hence, the rotation of the localized magnetic moments entails a redistribution of the free spins. In the materials with strong spin-orbit coupling such spin redistribution can create a charge current. However, in AFs the staggered localized moments create a staggered spin density. So, nonzero charge current can be created only in  crystals with a symmetry which allows for the field-like NSOT.    

To study the effect of inverse field-like NSOT we apply Onsager reciprocity relations to the set of thermodynamic fluxes $\{\dot{\mathbf{n}},\dot{\mathbf{m}}, \mathbf{j}\}$ and conjugated thermodynamic forces $\{\mathbf{H}_\mathbf{n}, \mathbf{H}_\mathbf{m}, \mathbf{E}\}$, where  $\mathbf{E}$ is the electric field (see Supplementary materials for derivation) and get expression for the current density:  
\begin{equation}\label{eq_SSGE}
\mathbf{j}=\frac{\lambda_\mathrm{NSOT}\sigma}{\gamma }\dot{\mathbf{n}}\times\hat{z}+\sigma \mathbf{E},
\end{equation}
where $\sigma$ is the electrical conductivity. 

From Eq.~(\ref{eq_SSGE}) it follows that any forced oscillations of the N\'eel vector induces an ac current with the frequency of oscillations. In particular, in the autooscillation regime $\dot{\mathbf{n}}=\Omega_\mathrm{dc}\hat{z}\times\mathbf{n}$  and rotations of the N\'eel vector  generate an ac current with the frequency $\Omega_\mathrm{dc}$ and the amplitude $j=\lambda_\mathrm{NSOT}\sigma M_sH_\mathrm{dc}/\alpha_\mathrm{eff}$. For $H_\mathrm{dc}=5.5$~mT ($j_\mathrm{dc}=2.75$~MA/cm$^2$) the frequency is 0.24 THz and $j=0.1$~MA/cm$^2$.  

The effect of inverse NSOT also contributes to the dissipation losses related with the in-plane rotation of the N\'eel vector (see Supplementary materials). For the considered in-plane geometry this contribution renormalises the effective damping constant $\alpha_\mathrm{eff}=\alpha_G+\lambda_\mathrm{NSOT}^2\sigma M_s/\gamma$. For CuMnAs this correction is $\sim 10^{-5}$ and can be neglected compared to the main contribution from the magnetic damping $\alpha_G$. 

{\it Conclusion --}
In summary, we have demonstrated that  a dc spin-current driven AF can be phase locked by an ac signal via the field-like NSOT. 
In the phase locking region the average magnetization of the sample is independent on the value of the dc spin current. 
The frequency of the incoming ac signal can be measured via precise measurements of the magnetization.  In addition, an AF  in autooscillation regime can generate an ac current and emit electromagnetic waves with the frequency of the spin-driven precession. Both effects can be used for tailoring narrow-band tunable detectors and emitters of THz radiation.

\begin{table}[h]\caption{Parameters of CuMnAs and Mn$_2$Au used for calculations}
		\begin{ruledtabular}
	\begin{tabular}{ccc}
		\hline 
		&  CuMnAs& Mn$_2$Au  \\ 
		\hline 
		$H_\mathrm{ex}$, T	\cite{Shick2010}& 200 &  1300\\ 
		\hline 
		$H_\mathrm{an}$, mT	\cite{Shick2010}& 5 &  7.5\\
		\hline 
		$\mu_0M_s$, mT\cite{Wadley2015a}	& 70 &  160\\ 
		\hline 
		$\omega_\mathrm{AFR}$, THz	&0.35  & 1.1 \\ 
		\hline 
		$\sigma$, S/cm	& 6250 & 4$\times10^4$\\ 
		\hline 
		$\lambda_\mathrm{NSOT}$\cite{Zelezny2014,Wadley2016}, s$^{-1}$A$^{-1}$cm$^2$&50&5\\
		\hline
		$\mu_0M_\mathrm{dyn}$, mT	& 2 &  0.7\\ 	
		\hline 
		$\alpha_G$ & 2$\times10^{-3}$ & $10^{-3}$ (estimate)\\ 	
		\hline 
		$\theta_H$ for Pt & 10~\% \cite{Sinova2015} &10~\% \cite{Sinova2015} \\ 	
		\hline 
		$H_\mathrm{dc}/j_\mathrm{dc}$ & 1 mT/(MA/cm$^2$) &1 mT/(MA/cm$^2$)  \\ 	
		\hline 
	\end{tabular} \label{Tab_data}
\end{ruledtabular}
\end{table}
We acknowledge the support from the Humboldt Foundation,   the ERC Synergy Grant SC2 (No. 610115), the EU FET Open RIA Grant no. 766566, the Collaborative Research Center (SFB/TRR) 173 SPIN+X,  from the Ministry of Education of the Czech Republic Grant No. LM2015087 and LNSM-LNSpin, and from the Grant Agency of the Czech Republic Grant no. 14-37427.
%

\appendix
\section{The inverse Ne\'eel spin orbit torque effect}
In this section we follow the Onsager reciprocity procedure of Ref. \onlinecite{Hals2011}.
In AFs with a  crystal structure like the one of CuMnAs and Mn$_2$Au  charge current with density $\mathbf{j}$ creates field-like torques \cite{Zelezny2014} $\mathbf{T}^\mathrm{A}=\lambda_\mathrm{NSOT}\mathbf{M}^\mathrm{A}\times\hat{z}\times\mathbf{j}$ and $\mathbf{T}^\mathrm{B}=-\lambda_\mathrm{NSOT}\mathbf{M}^\mathrm{B}\times\hat{z}\times\mathbf{j}$ on each of the magnetic sublattices A and B.
  This effect, which is similar to the inverse spin-galvanic effect \cite{Edelstein1990}, is called the N\'eel spin orbit torque (NSOT) effect. Physically it originates from the current-induced nonequilibrium distribution of the electrons with the certain spin polarization. In the inverse NSOT (similar to the spin-galvanic effect \cite{Ganichev2002}) the magnetic dynamics of an AF should induce an electrical current (field).

In order to find relation between the N\'eel vector and current density, we first represent the equations of motion for the magnetic sublattices in terms of the N\'eel vector $\mathbf{n}=(\mathbf{M}_\mathrm{A}-\mathbf{M}_\mathrm{B})$ and the magnetization vector $\mathbf{m}=(\mathbf{M}_\mathrm{A}+\mathbf{M}_\mathrm{B})$: 
\begin{eqnarray}\label{eq_dynamics}
\dot{\mathbf{n}}&=&\gamma\mathbf{H}_\mathbf{m}\times\mathbf{n}+\lambda_\mathrm{NSOT}\mathbf{m}\times(\hat{z}\times\mathbf{j}),\\
\dot{\mathbf{m}}&=&(\gamma\mathbf{H}_\mathbf{n}-\frac{\alpha_G}{2M_s}\dot{\mathbf{n}})\times\mathbf{n}+\lambda_\mathrm{NSOT}\mathbf{n}\times(\hat{z}\times\mathbf{j}),\nonumber
\end{eqnarray}
where $\mathbf{H}_\mathbf{m}=-\partial w/\partial \mathbf{m}$ and $\mathbf{H}_\mathbf{n}=-\partial w/\partial \mathbf{n}$  are the effective fields thermodynamically conjugated to $\mathbf{m}$ and $\mathbf{n}$, $w$ is the density of magnetic energy, $M_s$ is sublattice magnetization, $\alpha_G$ is the Gilbert damping  paramemters, $H_\mathrm{ex}$ parametrises intersublattice exchange in units of the magnetic field. In Eqs.~(\ref{eq_dynamics}) we have neglected the terms of the order of $\mathbf{m}^2$ because $m\ll1$. The last terms in the right hand side of Eqs.~(\ref{eq_dynamics}) correspond to the NSOT.

Then we apply the Onsager reciprocity relations to the set of thermodynamic fluxes $\{\dot{\mathbf{n}},\dot{\mathbf{m}}, \mathbf{j}\}$ and conjugated thermodynamic forces $\{\mathbf{H}_\mathbf{n}, \mathbf{H}_\mathbf{m}, \mathbf{E}\}$, where  $\mathbf{E}$ is the electric field.

Assuming that $\mathbf{j}=\sigma\mathbf{E}$, where $\sigma$ is conductivity, and using Eq.~(\ref{eq_dynamics}), we calculate the Onsager coefficients $\mathcal{L}_{\dot{\mathbf{n}},\mathbf{E}}$, $\mathcal{L}_{\dot{\mathbf{m}},\mathbf{E}}$ as follows:
\begin{eqnarray}\label{eq_Onsager_coefficients_direct}
\mathcal{L}_{\dot{\mathbf{n}},\mathbf{E}}&=&\lambda_\mathrm{NSOT}\sigma\left(\begin{array}{ccc}
-m_z& 0 & 0 \\ 
0 & -m_z & 0 \\ 
m_x & m_y & 0 \\ 
\end{array} \right),\nonumber\\
\mathcal{L}_{\dot{\mathbf{m}},\mathbf{E}}&=&\lambda_\mathrm{NSOT}\sigma\left(\begin{array}{ccc}
-n_z& 0 & 0 \\ 
0 & -n_z & 0 \\ 
n_x & n_y & 0 \\ 
\end{array} \right).
\end{eqnarray}
Magnetic vectors $\mathbf{m}$ and $\mathbf{n}$ change sign under the time reversion, while the electrical charge does not. This means that $\mathcal{L}_{\mathbf{E},\dot{\mathbf{n}}}=-\mathcal{L}_{\dot{\mathbf{n}},\mathbf{E}}^\mathrm{T}$ and  $\mathcal{L}_{\mathbf{E},\dot{\mathbf{m}}}=-\mathcal{L}_{\dot{\mathbf{m}},\mathbf{E}}^\mathrm{T}$. Thus, 
\begin{equation}\label{eq_Onsager_relation_inverse}
\mathbf{j}=\lambda_\mathrm{NSOT}\sigma\hat{z}\times\left(\mathbf{n}\times\mathbf{H}_\mathbf{m}+\mathbf{m}\times\mathbf{H}_\mathbf{n}\right)+\sigma\mathbf{E}.
\end{equation}
Taking into account that \cite{Gomonay2010} $\mathbf{H}_\mathbf{m}=-H_\mathrm{ex}\mathbf{m}/(2M_s)=\mathbf{n}\times\dot{\mathbf{n}}/4\gamma M_s^2$ and $\mathbf{n}^2=4M_s^2$, we get ultimately
\begin{equation}\label{eq_current_Onsager}
\mathbf{j}=\frac{\lambda_\mathrm{NSOT}\sigma}{\gamma }\dot{\mathbf{n}}\times\hat{z}+\sigma\mathbf{E} ,
\end{equation}
which coicides with Eq.~(6) of the main text.

 \section{Effective damping constant due to N\'eel spin-orbit torque}
According to Eq.~(\ref{eq_current_Onsager}), any oscillations of the N\'eel induce ac current $\mathbf{j}_\mathrm{ac}\propto \dot{\mathbf{n}}\times\hat{z}$. Substituting  the relation (\ref{eq_current_Onsager}) into equation of motion for the N\'eel vector (Eq.~(1) of the main text) we obtain equation of motion in the form
\begin{eqnarray}\label{eq_motion_antiferromagnet_with_newdamping}
&&\mathbf{n}\times\left[\ddot{\mathbf{n}}+2\gamma H_\mathrm{ex}\left(\underline{\alpha_G\dot{\mathbf{n}}+\frac{\lambda^2_\mathrm{NSOT}}{\gamma} \sigma M_s\hat{z}\times \dot{\mathbf{n}}\times\hat{z}}\right)\right.\nonumber\\&&\left.-2\gamma^2H_\mathrm{ex}M_s\mathbf{H}_\mathbf{n}\right]=\mathrm{external}\quad\mathrm{torques}
\end{eqnarray}
The underlined terms proportional to $\dot{\mathbf{n}}$ represent damping. For the in-plane rotation  ($\dot{\mathbf{n}}\perp \hat{z}$) the effective damping constant $\alpha_\mathrm{eff}=\alpha_G+\lambda_\mathrm{NSOT}^2\sigma M_s/\gamma$.

\end{document}